\documentclass[a4paper]{aa}

\usepackage{psfig}
\usepackage{times}

\def\simlt {\lower.5ex\hbox{$\; \buildrel < \over \sim \;$}}
\def\simgt{\lower.5ex\hbox{$\; \buildrel > \over \sim \;$}}

\def\NH {$N_{\rm H}$\ }

\begin{document}
   \thesaurus{ 08(08.06.1; 08.14.1; 09.09.1 RCW 86; 09.19.2; 13.25.5)}

    \title{An unresolved X-ray source inside the supernova remnant RCW 86}

    \author{Jacco Vink\inst{1} \and Fabrizio Bocchino\inst{2,3} \and Francesco Damiani\inst{3} \and Jelle S. Kaastra\inst{4}}

    \offprints{J. Vink (jvink@aip.de)}

    \institute{
	Astrophysikalisches Institut Potsdam,
	An der Sternwarte 16, D-14482 Potsdam, Germany
	\and
	Astrophysics Division, Space Science Department of ESA, ESTEC
	Postbus 299, 2200 AG Noordwijk, The Netherlands
	\and
	Osservatorio Astronomico di Palermo, Piazza del Parlamento 1, 
	I-90134 Palermo, Italy
	\and
	SRON, Laboratory for Space Research,
	Sorbonnelaan 2, NL-3584
        CA Utrecht, The Nether\-lands 
	}
\date{Received  21 June 2000 / Accepted 15 August 2000}

\maketitle

\begin{abstract}
We report on the discovery of an unresolved X-ray source inside the supernova 
remnant G315.4-2.3 (RCW 86). 
The source is located 7\arcmin\ to the Southwest of the 
geometrical centre and may be close to the actual explosion centre of the 
supernova, which makes this a candidate for the stellar remnant associated
with RCW 86.  
However, the presence of a possible optical counterpart with $V \sim 14$
at  3\arcsec\ from the X-ray position and evidence for long term 
variability means that the source is probably an active star.
A better X-ray position and better X-ray spectroscopy along with an 
identification of the optical source are needed to exclude the X-ray source as
a neutron star candidate.
\keywords{
Stars: flare -- Stars: neutron --
ISM: individual objects: RCW 86 -- ISM: supernova remnants -- X-rays: Stars }  
\end{abstract}

\section{Introduction}
In recent years it has become clear that young neutron stars
do not necessarily manifest themselves as radio pulsars.
Instead a large variety of unresolved objects associated with 
supernova remnants are thought to be the 
stellar remnants of the explosion (see Helfand \cite{Helfand98} for a review).
Examples are the point-like sources recently discovered in Cas A 
(Tananbaum \cite{Tananbaum}) 
and Puppis A (Petre et al. \cite{Petre96}), the enigmatic variable source in RCW 103 (Gotthelf et al. \cite{Gotthelf99a}), and a handful 
of relatively slow rotating X-ray pulsars called ``anomalous X-ray pulsars'' 
or AXPs (see Mereghetti~\cite{Mereghetti98} for a review).

Here we report on our analysis of an unresolved 
X-ray source in the supernova remnant RCW 86 (G315.4-2.3, MSH 14-6{\it 3}).
We discovered it during our work on the X-ray properties of the 
remnant (Vink et al. \cite{Vink97}, Bocchino et al. \cite{Bocchino2000}).
The source qualifies as the possible stellar remnant associated with
RCW 86, but the presence of a possible  optical counterpart and long term
source variability make this identification, as we will show, uncertain.

RCW 86 is the candidate remnant of the supernova AD 185 (Clark \& Stephenson
\cite{Clark},  Strom \cite{Strom94}), but the interpretation of the 
Chinese records is ambiguous (Chin \& Huang \cite{Chin}).
The large extent of the remnant (40\arcmin) can only be reconciled with
an explosion as recent as AD 185, if its distance does not exceed 1~kpc too 
much.
However, Rosado et al. (\cite{Rosado}) have pointed out that the kinematic
distance towards RCW 86  seems to be higher, namely  $2.8\pm0.4$~kpc.
In that case RCW 86 may be physically
associated with an OB association (Westerlund \cite{Westerlund}).
The presence of a stellar remnant in RCW 86 would establish the nature of the
supernova as a core collapse supernova (Type II or Ib/c), and therefore
a connection with the OB association would be more likely.

\begin{table}[t]
        \caption[]{A summary of ROSAT and Einstein observations which 
cover the point source. Dates refer to the start date of the observation.
\label{dates}}
{
        \begin{flushleft}
                \begin{tabular}{lllll}
\hline\noalign{\smallskip}
\noalign{\smallskip}
Date &  ID & Target & Instrument & Exposure\\
 d/m/y    &          &   &  & s \\
\noalign{\smallskip}\hline
\noalign{\smallskip}
 4/2/1980 & h1437s62 & SW & Einstein HRI &  8756\\
 4/9/1992 & rp500078n00 & NE  & ROSAT PSPC & 3212 \\
 2/2/1993 & rp500079n00 & SW  & ROSAT PSPC & 4567 \\
 3/2/1993 & rp500078a01 & NE  & ROSAT PSPC & 2224 \\
18/8/1993 & rp500078a02 & NE  & ROSAT PSPC & 4509 \\
 2/9/1994 & rh500077a01 & SW  & ROSAT HRI & 2447 \\
4/8/1995  & rh500077a02 & SW & ROSAT HRI & 6034\\
\noalign{\smallskip}\hline
                \end{tabular}
        \end{flushleft}
}
\end{table}

\begin{figure}
\psfig{figure=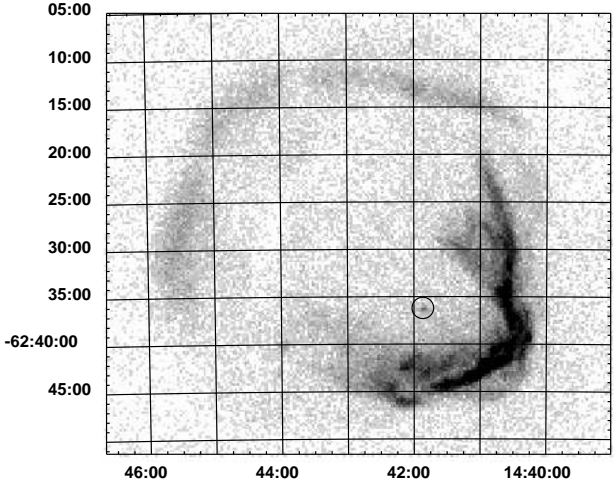,width=8.5cm}
\caption[]{RCW 86 as observed by the PSPC 
(SW pointing). The circle indicates the position of the point source.
\label{maxlik}}
\end{figure}

\begin{table}[tb]
        \caption[]{Best fit positions and count rates 
of the point source. 
The first three rows list the result of
maximum likelihood fits to the ROSAT and Einstein HRI data.
The statistic $-2\ln\lambda$ has a $\chi^2$ distribution with three degrees of
freedom. Position errors do not include systematic errors and
correspond $\Delta(-2\ln\lambda) = 4.6$ (90\% confidence regions).
The PSPC positions and count rates were estimated with a wavelet 
analysis method (Damiani et al. \cite{Damiani}) 
using the energy channels 20-200 ($\sim 0.2-2$~keV).
\label{positions}}
{\scriptsize
        \begin{flushleft}
                \begin{tabular}{llllll}
\hline\noalign{\smallskip}
 ID   &  RA (J2000) & DEC (J000) & error & count rate &  $-2\ln\lambda$\\
      &  \ h \ \ m \ \ \ s   &\ \degr\ \ \ \ \arcmin\ \ \ \ \arcsec & \arcsec & $10^{-3}$cnts/s\\
\noalign{\smallskip}\hline
\noalign{\smallskip}
h1437s62    & 14 41 51.52 & -62 36 13.1 & 2.6 
& $1.2\pm$ 0.4 & 28.6 \\
rh500077a01 & 14 41 51.64 & -62 36 11.9 & 3.5 
&$2.1\pm1.1$ &12.4\\
rh500077a02 & 14 41 51.11 & -62 36 13.2 & 1.9 
&$3.4\pm0.8$ & 55.3\\

rp500078n00 & 14 41 53.3 & -62  36 37 & 65 & $10.9 \pm3.9$\\
rp500079n00 & 14 41 51.4 & -62 36 8 & 15  & $17.9\pm2.7$\\
rp500078a01 & 14 41 49.0 & -62 35 52 & 59 & $19.4\pm6.5$\\
rp500078a02 & 14 41 48.9 & -62 36 22 & 39 & $26.8\pm6.8$\\

\noalign{\smallskip}\hline
                \end{tabular}
        \end{flushleft}
}
\end{table}

\section{Data analysis}
Our analysis is based on archival ROSAT PSPC and HRI data and on
an observation of RCW86 by the Einstein HRI. 
Both HRI instruments have a similar spatial resolution of 4\arcsec\ FWHM, 
but the ROSAT HRI is more sensitive.
The PSPC has a resolution of only 30\arcsec\ FWHM,
is more sensitive than the HRI and has better spectral capabilities 
covering the energy range of 0.1 to 2.4~keV with a spectral 
resolution $\Delta E/E = 0.4$ at 1~keV.
The data used for our analysis are summarised in Table~\ref{dates}.
The starting point of the analysis are the basic screened event
lists. Further processing, like photon extraction and barycentric correction,
was done with NASA's {\it ftools} v4.2 package.

\begin{figure}[htb]
\psfig{figure=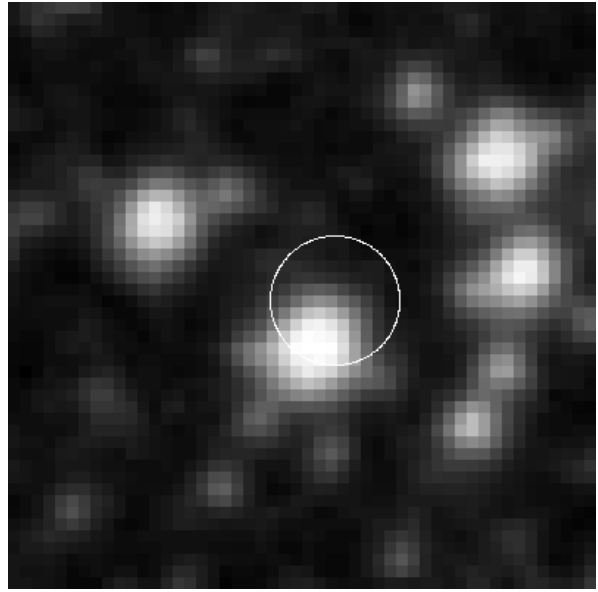,width=7.8cm}
\caption[]{
A 50\arcsec\  $\times$\ 50\arcsec\ field 
image taken from the Digital Sky Survey 2.
Overplotted is a circle with a radius of 5\arcsec\ centered on 
the point source and roughly corresponding to a 95\% confidence region.
\label{optical}}
\end{figure}

\subsection{Position of the source}
In order to find an accurate position for the unresolved source
we used the HRI data and
fitted the point spread function to the source using the very sensitive
maximum likelihood fits 
(Cash \cite{Cash} and e.g. Hasinger et al. \cite{Hasinger94}). 
We used a field of 
80\arcsec $\times$ 80\arcsec\ roughly centred on the source.
The normalisation was coupled to the background level using
$N_{\rm total} = N_{\rm bgd} + N_{\rm src}$, 
$N_{\rm total}$ being the observed number of photons and  
$N_{\rm bgd}$, $N_{\rm src}$ the fitted background and source 
normalisations.
The results of these maximum likelihood fits are listed in 
Table~\ref{positions}, which also lists 
the source parameters
for the PSPC data as found with a wavelet analysis 
(Damiani et al. \cite{Damiani}).
One of the HRI detections has a significance level 
of $2.7\sigma$, not
good enough to claim a detection by itself, but combined with the
other detections the source parameters can be regarded as meaningful.

Unfortunately, positions based on ROSAT sometimes suffer from errors in the
attitude calculations which are typically 6\arcsec\ 
(Hasinger et al. \cite{Hasinger94} and private communication). 
The observed scatter in the positions based on the Einstein and ROSAT HRI
images, as compared to the statistical position errors,
suggests that also our results are affected by systematic errors.
Einstein HRI observations are less affected by systematic position errors,
the typical systematic position error being $\sim$ 2\arcsec\ 
(Van Speybroeck et al. \cite{speybroeck}).
Adding statistical and systematic errors in quadrature the
weighted average of the positions based on HRI images is
$\alpha = $14h 41m 51.42s and 
$\delta = -62$\degr36\arcmin 12.9\arcsec\ (J2000) with a $1\sigma$ position
error of approximately 3\arcsec. For a two dimensional gaussian this 
translates into a 95\% confidence radius of 5\arcsec.
Note that a very bright, unresolved, radio source inside the remnant with 
coordinates $\alpha = 14^h41^m44.5^s$ and 
$\delta = -62$\degr34\arcmin 47\arcsec\ (J2000)
is clearly not associated with the unresolved X-ray source 
(Dickel et al. \cite{Dickel}).

\begin{table*}
        \caption[]{Results of the spectral fits to the PSPC data.
The luminosities are normalised to a distance of 1~kpc.
For each model the spectrum was 
fitted with \NH\ $= 1.7\ 10^{21}$~cm$^{-2}$ and with \NH\ as a free parameter.
Error ranges correspond to $\Delta \chi^2 = 2.7$, or 90\% confidence limits.
\label{spectral}}
        \begin{flushleft}
                \begin{tabular}{llllll}
\hline\noalign{\smallskip}
Model & Parameter & Normalization &\NH & $L_X(0.2-2.5\,{\rm keV})$& $\chi^2$/d.o.f.\\
& & & $ 10^{21}$~cm$^{-2}$ & $10^{31}$~erg/s \\
\noalign{\smallskip}\hline\noalign{\smallskip}
Blackbody  & k$T = 0.17 \pm 0.04$~keV & 
$(4.6^{+16.8}_{-3.5})\ 10^{10}$~cm$^2$ & 1.7 & 3.6 & 8.4/9\\
           &  k$T = 0.23 \pm 0.06$~keV &  
$(0.59^{+1.9}_{-0.35})\ 10^{10}$~cm$^2$ & $0.12^{+4.0}_{-0.2}$ & 1.6 & 1.1/8\\
\noalign{\smallskip}
Power Law  & $\Gamma = 3.9^{+0.12}_{-0.9}$ & $(9.7\pm 2.7)\ 10^{39}$~ph/s/keV @ 1~keV&
1.7 & 16.9 & 4.1/9\\
	   & $\Gamma = 2.9\pm 1.0$ & $(7.5\pm2.9)\ 10^{39}$~ph/s/keV @ 1~keV &
$0.84\pm 0.47$ & 5.1 & 1.7/8\\
\noalign{\smallskip}
hot thin plasma &  k$T = 0.68 \pm 0.41$~keV & $(6.8 \pm 2.4)\ 10^{53}$~cm$^{-3}$ & 1.7 & 1.9 & 14.7/9\\	   
\ \ \ \ \ (mekal)  &   k$T = 1.12^{+7.1}_{-0.57}$~keV & $(6.6^{+1.4}_{-0.3})\ 10^{53}$~cm$^{-3}$ & $< 0.3$ & 1.3 & 4.3/8 \\
\noalign{\smallskip}
idem,$Z=0.1$ &  k$T = 0.63 \pm 0.35$~keV & $(4.4^{+4.1}_{-1.2})\ 10^{54}$~cm$^{-3}$ & 1.7 & 3.0 & 10.5/9 \\
                  &  k$T = 0.87^{+0.96}_{-0.33}$~keV & $(2.6 \pm 0.7)\ 10^{54}$~cm$^{-3}$ & $0.3 \pm 0.2$ & 1.9 & 1.0/8\\
\noalign{\smallskip}\hline
                \end{tabular}
        \end{flushleft}
\end{table*}

\subsection{Spectral analysis}
For the spectral analysis of the PSPC data of the unresolved source
we extracted photons using a circular area with a radius of
32\arcsec\ for the SW (on-axis) pointing and 44\arcsec\ for the other 
pointings. We estimate that with such radii we cover roughly 90\% of the 
photons coming from the unresolved source 
(c.f. Hasinger et al. \cite{Hasinger92}). 
Background spectra, extracted
from an annulus around the source, were appropriately scaled and subtracted
from the source spectra. 
The combined spectrum consists of 177 net source counts. The spectrum
was rebinned in order to have at least 15 counts per bin.

For our spectral analysis we used he spectral fitting program SPEX\footnote{The black body model in SPEX v1.10 contains a small bug which we fixed for this analysis. The thin thermal plasma or CIE (collisional ionization equilibration) model is similar to the {\em mekal} model in xspec.} (Kaastra et al. \cite{Kaastra96}).
Since we want to know whether the source qualifies as the potential stellar
remnant associated with RCW 86, 
we fitted the spectrum with several emission models both with the 
interstellar absorption value fixed at \NH\ $= 1.7\, 10^{-21}$~cm$^{-2}$,
the typical absorption value for the X-ray emission of the supernova remnant 
(Vink et al. \cite{Vink97}),
and with \NH\ as an additional free parameter. 
The results are listed in Table~\ref{spectral}.
The best fit values of \NH\ for all models seem to be in favor of a low
absorption column towards the source, but also models with fixed
\NH\ give acceptable reduced $\chi^2$ values. 
The fact that models with three parameters result in very low 
reduced $\chi^2$ values (i.e. far from the  $\chi^2$ expectation value),
suggests that the statistics of the data is not really good enough to fit
models with three or more parameters.
All models provide reasonable fits to the data with only the thin plasma
model with solar abundances and fixed $N_{\rm H}$ having a reduced $\chi^2$ 
substantially larger than 1. 
The spectrum appears to be rather soft as indicated by
the steep power law index, $\Gamma$, and the low black body temperature.

\subsection{Timing analysis}
We searched the four PSPC observations for possible pulsations using the
Rayleigh method (Buccheri et al. \cite{Buccheri}). 
This method is one of the most sensitive 
methods and it does not involve any binning of the data. A sensitive method
is needed as the longest PSPC observations yielded only 101 events.
We searched in each set for pulsations in the period range 0.02 to 300 s,
sampling the frequency range with step of $1/T_{\rm obs}$ 
with $T_{\rm obs}$       
the total length of the observation.
We compared the periodograms to look for peaks showing up in two or more 
periodograms at or near the same period.
Such correlations were, however, not found.
The peak values of the Rayleigh statistic,
$Z^2_1 \sim 28$, imply an upper limit to the pulsed fraction of 
$\sim $20\%.

As for the variability on the timescales of month, at first
sight there is little evidence for variability as all measured PSPC count 
rates are consistent with a count rate of $(17.0\pm2.8)\ 10^{-3}$
cnts/s. However, if we convert the Einstein and ROSAT HRI count rates 
to PSPC count rates using the best fit power law model in Table~\ref{spectral}
(the conversion factors are 4.9 and 2.7, respectively) we get 
the following PSPC count rates (in the same order as in Table~\ref{positions}):
$(5.8\pm 1.9)\ 10^{-3}$ cnts/s, 
$(5.7\pm 2.7)\ 10^{-3}$ cnts/s,
and $(9.0\pm 2.1)\ 10^{-3}$ cnts/s.
The dependence of the ROSAT HRI/PSPC conversion factor on the chosen
model is small (9\%), but the conversion factor 
for the Einstein HRI count rates is more model dependent, 
varying from 4.6 to 8.0.
Even taking into account the model uncertainties it is
clear that the observations are not consistent with a constant source count 
rate, although it is a strange coincidence that low count rates were 
only observed by the HRI instruments. 
Source contamination with the PSPC instrument
seems unlikely, as no other unresolved sources are seen with the HRI 
instruments near the point source.
Therefore, the X-ray source
is very likely variable on a time scale of months to years.

\begin{figure}
\psfig{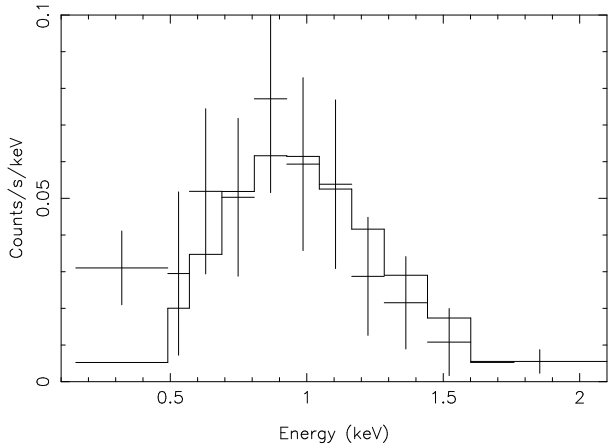}
\caption[]{
The PSPC spectrum of the unresolved source in RCW 86. It consists of
a combination of 4 individual observations. The solid line shows the
best fit black body model with \NH\ fixed to a value of $1.7\ 10^{21}$~cm$^{-2}$.
\label{spectrum}}
\end{figure}

\section{Discussion}
So, could this X-ray source be the stellar remnant associated with RCW 86?
The radius of the star, as inferred from the black body fit, is
1.7~km at a distance of 2.8~kpc. This is too small for a neutron star,
but the spectrum may not be a black body. 
The X-ray luminosity is lower than the luminosity of typical AXPs,
but it is consistent with the surface luminosity of young neutron stars
in case pion cooling is important (Umeda et al. \cite{Umeda}).
Also emission from a black hole accreting supernova fall back material
should be considered as an alternative model (Umeda et al. \cite{Umeda},
Chakrabarty et al. \cite{Chakrabarty}).
The softness of the X-ray spectrum is a property shared with the 
unresolved sources in Puppis A,  G296.5+10.0 and AXPs
(Mereghetti et al. \cite{Mereghetti96}).
Its long term variability is more in line with the behaviour of
an active star, but a well established neutron star candidate like 
the point 
source in RCW 103 is also variable on similar time scales 
(Gotthelf et al. \cite{Gotthelf99a}).

The position of the X-ray source is roughly at an angular distance of 7\arcmin\
from the geometrical center of the supernova remnant. This corresponds
to a transverse velocity \simlt 1200~km/s, if the supernova
remnant is at a distance of 2.5~kpc and 5000~yr old, 
or if the distance is 1~kpc and an age of 1800~yr.
Such a kick velocity is rather high, but still consistent with observations
of other neutron stars (Lyne \& Lorimer \cite{Lyne}).
Note, however, that the remnant
is far from circularly symmetric, and
the contrast in emission between the Northeast and Southwest of the remnant
(see e.g. Vink et al. \cite{Vink97}, Bocchino et al. \cite{Bocchino2000}) 
suggests that in the Southwest the 
shock wave is encountering a denser medium, which could mean that the actual
explosion center was more to the Southwest of the geometrical center and
closer to the unresolved source.
Interestingly, the remnant is quite symmetric in
the Northeast/Southwest direction and the X-ray source is roughly located on
the axis of symmetry.
The position of the point source is one of its salient properties.
The wavelet detection code has also detected
another significant point source inside the remnant, 7\arcmin\
from the Northern shell with coordinates 
$\alpha = $14h 43m 47.0s and $\delta =$ -62\degr 19\arcmin 29\arcsec.
The source seems embedded  in a small 1\arcmin\ size extended structure.
However, spectral analysis of this source rules out models with
non-thermal or black-body radiation and the spectrum is consistent with 
thermal X-ray emission, characteristic for the North rim as
reported by Bocchino et al. (\cite{Bocchino2000}). 
This suggests that this a small fragment of the shell seen in 
projection.

It seems unlikely that the unresolved X-ray source is much further away
than RCW 86, as the X-ray absorption column is too low
for a distant source near the galactic plane.
This means that it is unlikely that it is either an AGN or
an X-ray binary. For the source to be a typical X-ray binary with a 
luminosity  of $10^{36}-10^{38}$~erg/s,
it should be at a distance of 300~kpc. This is incompatible with both the 
X-ray absorption and with the size of our galaxy.

The main argument against an unambiguous identification of the source 
with a stellar remnant is the presence of a candidate optical counterpart.
The USNO-A2.0 catalogue (Monet et al. \cite{Monet}) contains one
source at an angular distance of 3\arcsec\ to the unresolved X-ray source 
(see Fig. \ref{optical}).
The star has magnitudes $r=13.6$ and $b=15.9$, which suggest a K-star, 
although the magnitude errors and the unknown extinction make 
a late type G-star or a early type M star also possible 
(Zombeck \cite{Zombeck}). 
The typical V magnitude of this star would be $V \sim 14$,
which, for a main sequence star, implies a distance of $\sim 250$~pc. 
For this distance the X-ray luminosity of the X-ray source
is $L_X  \simgt  6\ 10^{29}$. This is consistent with the X-ray luminosity
of a bright active star (see e.g. Zombeck \cite{Zombeck} p. 88, 
Agrawal et al. \cite{Agrawal}).
The possibility that this is just a chance alignment can be estimated using
the USNO-A2.0 catalogue. There are about  0.0015 stars per square arcsec 
with $m_{\rm R} \leq 16$ in the field around the X-ray source,
which means that the chance to find a bright star within the 95\% error circle 
(see Fig.~\ref{optical}) is $\sim$11\%.

The evidence for long term variability
and the presence of a late type star close to the X-ray position
favour the identification of the X-ray source with an active star,
rather than with the stellar remnant associated with RCW 86.
However, to firmly establish this identification,
optical spectroscopy of the possible optical counterpart and
X-ray observations with a better positional accuracy of the source are needed.
Improved X-ray spectroscopic data can also be used to distinguish 
between the typical optically thin thermal spectrum of an active star and 
the featureless spectrum of a neutron star. 

\begin{acknowledgement}
We thank Guenther Hasinger and Rick Harnden for their information
regarding the systematic position errors of ROSAT and Einstein. We thank
Richard Strom and John Dickel for discussions on the radio map of 
RCW 86/G315.4-2.3.
\end{acknowledgement}

\end{document}